%% file: main.tex
\documentclass[10pt, conference]{IEEEtran}

\usepackage[utf8]{inputenc}
\usepackage[margin=2.5cm]{geometry}
\usepackage{graphicx}
\usepackage{float}
\usepackage{tabularx}
\usepackage{amsmath}
\usepackage{amssymb}
\usepackage{amsthm}
\usepackage[dvipsnames]{xcolor}
\usepackage[hyphens]{url}
\usepackage{hyperref}
\usepackage{cleveref}
\usepackage{bm}
\usepackage{gensymb}
\usepackage{titlesec}
\usepackage{subcaption}
\usepackage[ruled,vlined]{algorithm2e}
\usepackage{stmaryrd}
\usepackage{enumitem}
\usepackage{float}

\newcommand{\hh}{\hspace{-2pt}}

\usepackage{orcidlink}

\usepackage[style=ieee, maxnames=3, maxbibnames=3, backend=bibtex]{biblatex}
\hypersetup{
    colorlinks,
    citecolor=black,
    filecolor=black,
    linkcolor=black,
    urlcolor=black
}

\usepackage[colorinlistoftodos]{todonotes}

\begin{document}
\IEEEoverridecommandlockouts

\title{On the challenges of using D-Wave computers to sample Boltzmann Random Variables}

\author{\IEEEauthorblockN{Thomas Pochart}
\IEEEauthorblockA{\textit{EDF Lab Saclay}\\
\orcidlink{0000-0003-4368-7239} {0000-0003-4368-7239}}
\and
\IEEEauthorblockN{Paulin Jacquot}
\IEEEauthorblockA{\textit{EDF Lab Saclay}\\
\orcidlink{0000-0002-1086-2494} {0000-0002-1086-2494}}
\and
\IEEEauthorblockN{Joseph Mikael}
\IEEEauthorblockA{\textit{EDF Lab Saclay, FiME Laboratory} \\
\orcidlink{0000-0002-0416-5297} {0000-0002-0416-5297}}

\thanks{This research is supported by the departement Optimization, SImulation,
RIsk and Statistics for energy markets (OSIRIS) of EDF Lab and FiME (Finance des marchés de l'energie) Laboratory which are gratefully acknowledged.}
}


\maketitle


\begin{abstract}
Sampling random variables following a Boltzmann distribution is an NP-hard problem involved in various applications such as training of \textit{Boltzmann machines}, a specific kind of neural network. 
Several attempts have been made to use a D-Wave quantum computer to sample such a distribution, as this could lead to significant speedup in these applications.
Yet, at present, several challenges remain to efficiently perform such sampling. We detail the various obstacles and explain the remaining difficulties in solving the sampling problem on a D-wave machine.
\end{abstract}

\noindent 

\input{introduction}
\input{challenges}
\input{conclusion}

\AtNextBibliography{\small}
\printbibliography
\end{document}

%% file: introduction.tex
\section*{Introduction}
Sampling Boltzmann-distributed variables associated with an Ising model is an NP-hard problem \cite{barahona1982computational}, which is notably involved in the training of the so-called \textit{Boltzmann machines} (BM), a particular kind of discrete neural network. Sampling Boltzmann distributions is classically performed using Monte-Carlo algorithms such as Gibbs sampling \cite{hinton2002training}, which remains time-consuming. 

Adiabatic quantum computing (AQC) allows in principle to efficiently find the minimal energy states of a user-defined Hamiltonian energy function. 
D-Wave \cite{headquarters2020technical} is a world leader on the implementation of AQC hardware, currently selling computation time on  machines with 5,000 qubits. 
While D-Wave machines were designed to be solvers and not samplers, the operating principle of AQC suggests that those machines could be efficiently used to sample from Boltzmann distributions.
In fact, Boltzmann sampling using a D-wave machine has already been investigated in the existing literature 
 ~\cite{dixit2021training}, ~\cite{liu2018adiabatic},~\cite{crawford2016reinforcement}, where the authors aim to train Boltzmann Machines.
Yet, the implemented protocol allowing to perform such sampling is usually difficult to reproduce.

In this paper, based on a review of the existing literature on the subject and our own numerical experiments with D-Wave machines, we describe the challenges and difficulties that exist in using available AQC hardware for Boltzmann sampling.
Although this paper does not bring a solution to the Boltzmann sampling problem, it contributes to identify why efficiently solving the Boltzmann sampling problem using AQC is, at least at the current stage and to our limited knowledge and technology access, out of reach.


%% file: challenges.tex
\newcommand{\Eising}{E_{_\mathrm{ising}}}

\section{Boltzmann Sampling using AQC}
\label{sec:bs}
\subsection{Problem statement}

\noindent Consider an Ising energy $\Eising$, i.e. a function of a state vector 
%
$\bm{S} \in \{ -1, 1\}^N$ of dimension $N\in\mathbb{N}$:
\begin{equation}
    \Eising(\bm{S}) = \sum\limits_{i=1}^N h_i S_i + \sum\limits_{i=1}^N \sum\limits_{j=i+1}^N J_{i,j} S_i S_j \: (+ \:  cst) \label{eq:EIS}
\end{equation}
to which we associate the Boltzmann distribution:  
\begin{equation}
    \mathbb{P}(\bm{S}) = \tfrac{1}{Z}{}e^{-\Eising(\bm{S})}, \   Z = \sum\limits_{\bm{x}} e^{-\Eising(\bm{S})} \label{eq:defBoltzmDist}
\end{equation}
where $Z$ denotes the associated partition function.

The Boltzmann sampling problem is thus, given an integer $K\geq 1$, to obtain $K$ sample vectors $(\bm{S}_k)_{k=1}^K$ that are distributed following \eqref{eq:defBoltzmDist}.

\subsection{Overview of computation in a D-Wave machine} \label{overviewdw}

\noindent A D-Wave machine finds the ground states of a given Ising energy  (minimizing $\Eising$ in \eqref{eq:EIS}) by:

\begin{enumerate}
    \item Applying to $N$ qubits a Hamiltonian:
        \begin{equation}
            \hat{\mathcal{H}}_{i} \propto  \textstyle\sum_{i} \hat{\sigma}^{(i)}_x \label{eq:Hi}
        \end{equation}
    \noindent where $\hat{\sigma}_x^{(i)}$ is the $x$ Pauli matrix applied to qubit $i$.
    \item Getting the set of qubits in its ground state.
    \item Applying between $t\hh=0$ and $t\hh=t_f$ the Hamiltonian:
        \begin{multline}
            \hat{\mathcal{H}}_{QA}(s) = -\frac{A(s)}{2} \left( \sum\limits_{i} \hat{\sigma}^{(i)}_x \right) + \\ \frac{B(s)}{2} \left( \sum\limits_{i} h_i \hat{\sigma}^{(i)}_z + \sum\limits_{i<j} J_{i,j} \hat{\sigma}^{(i)}_z \hat{\sigma}^{(j)}_z \right) \label{eq:HQAs}
        \end{multline}
        \noindent where
            $s=s(t)=t/t_f$ is a dimensionless ratio increasing linearly with time from $0$ to $1$, and 
  $A(s)$ and $B(s)$ are positive quantities homogeneous to energies such that $A(0)\hh\gg\hh B(0)$ and $A(1)\hh \ll\hh B(1)$.
    \item Measuring the states of the qubits, which are according to the adiabatic theorem very likely to be in a ground state of:
        \begin{equation}
            \hat{\mathcal{H}}_{Ising} \propto  \sum\limits_{i} h_i \hat{\sigma}^{(i)}_z + \sum\limits_{i<j} J_{i,j} \hat{\sigma}^{(i)}_z \hat{\sigma}^{(j)}_z \label{eq:Hf}
        \end{equation}
    \item Deducing a minimum of $\Eising$ using (the affine tranformation $x\mapsto 2x-1$ is used to switch from $\bm{x}\in\{0\,1\}^N$ to the Ising state vector $\bm{S}\in\{-1,1\}^N)$:
        \begin{multline}
            \forall \bm{x} \in \{0,1\}^N, \hat{\mathcal{H}}_{Ising} \left| x_1 \right\rangle \otimes \dots \otimes \left| x_N \right\rangle \\  \propto \Eising(2\bm{x}-\bm{1}) \left| x_1 \right\rangle \otimes \dots \otimes \left| x_N \right\rangle. \label{eq:HIsingEig}
        \end{multline}
    \item Repeating the previous steps to explore all solutions and compensate for failures.
\end{enumerate}


\subsection{Why an AQC 
could solve the Sampling problem}

\noindent Let $E(\bm{h},\bm{J},\bm{S})$ be the physical energy of the set of $N$ qubits when in configuration $\bm{S}$. When the system is in thermodynamic equilibrium at temperature $T$, its probability distribution states is a Boltzmann distribution given by:
\begin{equation}\label{eq:physicalBoltzmDist}
    \mathbb{P}(\bm{S}) = \tfrac{1}{Z}e^{-\frac{E(\bm{h},\bm{J},\bm{S})}{k_B T}} \ , \  Z = \textstyle\sum e^{-E(\bm{h},\bm{J},\bm{S})}. 
\end{equation}
where $Z$ is called the partition function. 
Thus, by giving a system of $N$ qubits a Hamiltonian 
\begin{equation}
    \hat{\mathcal{H}}_{Ising} = k_B T \left( \sum\limits_{i} h_i \hat{\sigma}^{(i)}_z + \sum\limits_{i<j} J_{i,j} \hat{\sigma}^{(i)}_z \hat{\sigma}^{(j)}_z \right) \label{eq:HfkBT}
\end{equation}
and putting this system in thermal equilibrium, one could determine experimentally Eq.\eqref{eq:defBoltzmDist} by sampling.


\section{Challenges for Effective AQC Sampling}
\label{sec:challenges}
\subsection{Challenge \#1. Eligible problems} \label{eligpb}
\noindent For physical reasons, in a D-Wave computer not every pair of qubits can be coupled together: the Ising Hamiltonian functions~\eqref{eq:Hf} that can effectively be implemented on D-Wave hardware have some $J_{i,j}$ that must be zero (qubits $i$ and $j$ not coupled). 
To overcome this limit, D-Wave relies on ``embedding'': a \textit{logical} qubit (from the mathematical problem) may be represented not by one \textit{physical} qubit, but by a chain of physical qubits under the constraint that those qubits must have the same value.\\
In practice, two (or more) qubits $i$ and $j$ can be statistically forced to take the same value in the ground state of an Hamiltonian by adding it $\gamma \hat{\sigma}_z^{(i)} \hat{\sigma}_z^{(j)}$, where $\gamma > 0$ is called the \textit{chain strength}. 
However, embedding has two major drawbacks:
\begin{itemize}[wide]
    \item It is not obvious to choose $\gamma$. The bigger this factor, the lesser the proportion of samples with broken chains, and therefore the less amount of samples required to sample from the target distribution through conditional probability.
    But as a scaling of the problem is performed before it is implemented and solved (see~\ref{esta}),  choosing a too large $\gamma$ yields the risk of scaling down the energies of multiple states so small 
    that they cannot be distinguished by the sensors.
    \item The embedding problem
    is known to be NP-hard~\cite{robertson1995graph}. As this problem is solved by the classical computer that controls the electronics of the quantum computer, this can seriously hinder the ability D-Wave has to efficiently solve problems. 
    However, it can be experimentally observed that on most concrete applications, the runtime advantage induced by the use of a D-Wave computer is still consequent \cite{adachi2015application},~\cite{willsch2020support},~\cite{chapuis2019finding}.
\end{itemize}

\subsection{Challenge \#2. Processor imperfections}

\noindent This part is mostly a summary of the relevant information found in D-Wave's technical documentation~\cite{headquarters2020technical}.

\smallskip

\subsubsection{Integrated Control Errors (ICEs)} \label{ICEs}

\noindent ICEs refer to the fact that when the user specifies the Ising energy \eqref{eq:EIS},
 the computer actually implements:
\begin{equation*}
    E^{\delta}_{Ising}(\bm{S}) = \sum\limits_{i=1}^N (h_i +\delta_{h_i}) S_i + \sum\limits_{i=1}^N \sum\limits_{j=i+1}^N (J_{i,j} + \delta_{J_{i,j}}) S_i S_j
\end{equation*}
where $\delta_\alpha$ denotes a small perturbation to a quantity $\alpha$.
There are five main causes for ICEs:
\begin{itemize}[wide]
    \item \textit{Background susceptibility:} The qubits are controlled with magnetic fields. These fields may also interact with their material background, and thus be perturbed.
    \item \textit{Flux noise of the qubits:} The qubits actually interact with the flux of the magnetic field, and are thus sensitive to the latter's fluctuations. 
    \item \textit{DAC quantization:} Digital Analog Converters (DACs) used to transmit an input signal to the electronics controlling qubits have a finite quantization step size.
    \item \textit{I/O System effects:} The annealing process involves time-dependent signals. As they are  delivered through I/O electronics that only have a finite bandwidth,  there can be signal distortion.
    \item \textit{Distribution of $\bm{h}$ scale across qubits:} Like any other device, the qubits aren't perfect replicas of each other and small differences between them exist.
\end{itemize}

\subsubsection{High-energy photon flux} \label{hepf}

\noindent The adiabatic theorem applies to isolated systems. It is not exactly the case of the set of qubits in the QPU: they interact not only with the magnetic field generated by the computer to control them, but also with radiations corresponding to a flux of relatively high energy photons that the system cannot be isolated from. These radiations tend to supply energy to the qubits, creating a risk of kicking them out of a ground state and hence taking them to an excited state. This risk grows with $t_f$, as the longer the anneal, the more the qubits get irradiated.

\smallskip

\subsubsection{Spin-bath polarization effect} \label{sbpe}

\noindent During annealing, the body of a qubit can acquire an undesired 
magnetic polarization, affecting its response to the magnetic field. Possible consequences are:
\begin{itemize} [wide]
    \item Contributions to background susceptibility yielding more ICEs. The induced ICE does not only depend of the current problem, but on the whole history of the qubit and thus possibly on the problems it previously solved.
    \item  The production of sample-to-sample correlations, i.e. bias of the QPU towards previously achieved configurations---including configurations reached when solving another problem.
\end{itemize}

\noindent The only solution to avoid this problem is to give time to the QPU to depolarize after being operated. However, this comes at the cost of an increased runtime, while the primary purpose of the QPU is to solve problems fast.

\smallskip

\subsubsection{Readout fidelity}

\noindent The reading step isn't perfect either. However, the readout fidelity, i.e. the proportion of correctly read qubits, is approximately of $99\%$~\cite{headquarters2020technical}; thus, this phenomenon is most of the time negligible.

\subsection{Challenge \#3. Reaching thermal equilibrium} \label{eqhack}
\noindent D-Wave computers are devices that fundamentally implement an adiabatic evolution meant to favor the ground states above all the others, thus escaping the equilibrium distribution described in~\eqref{eq:physicalBoltzmDist}. So not only it is natural for our target distribution to not be observed in a sample outputted by a D-Wave computer, but also consistently making such an observation would require one to go against the fundamental mechanics of the computer.

\smallskip

\noindent D-Wave computers let their users redefine up to a certain extent $s(t)$ (see~\ref{overviewdw}), allowing in particular for ``pauses'' in the annealing schedule. As Boltzmann distributions are to be reached in equilibrium setting, one may hope that such a pause would give the system enough time to reach such an equilibrium. This direction is explored in \cite{nelson2021high}, where authors  experimentally determine on simple problems which pauses yield the results that are closer to an actual Boltzmann distribution. 
Yet despite the theoretical and experimental insights provided, the road towards Boltzmann sampling is still full of pitfalls as:

\begin{itemize}[wide]
\item upon experimentally determining which pause of the anneal schedule yields a sampleset that is the closest to a Boltzmann distribution, authors of \cite{nelson2021high} retrofit an Ising model corresponding to this distribution. Hence, even if a pause-based protocol could yield a Boltzmann distribution, it would \textit{a priori} not be the target distribution (thus not solving the sampling problem);
\item the framework of parameters that is considered is restrictive: the weights of the input Ising model must belong to $\{-1, 0, 1\}$, and it must not feature more than 16 qubits. For such parameters, a classical brute-force algorithm can estimate the target distribution.
\end{itemize}

\subsection{Challenge \#4. Hamiltonian scaling} \label{Hscal}

\paragraph{Problem statement}
\noindent There are two fundamental differences between the target distribution~\eqref{eq:defBoltzmDist} and the physical distribution~\eqref{eq:physicalBoltzmDist}:

\begin{itemize}[wide]
\item The physical distribution involves a factor $1/k_BT$ that is missing from the target distribution.
\item The $E(\bm{h},\bm{J},\bm{S})$ of the physical distribution~\eqref{eq:physicalBoltzmDist} are not exactly the $\Eising(\bm{S})$ of the target distribution~\eqref{eq:defBoltzmDist}: while the former are physical energies, that can be expressed in Joules, the latter are dimensionless numerical quantities, that must be expressed without units. Therefore, given a Ising Hamiltonian with weights ($\bm{h}$,$\bm{J}$) and a state $\bm{S}$, there exists some physical energy $E_0(\bm{h},\bm{J},\bm{S})$ such that $E(\bm{h},\bm{J},\bm{S}) = \Eising(\bm{S})E_0(\bm{h},\bm{J},\bm{x})$.
\end{itemize}
\noindent While the target distribution is defined in \eqref{eq:defBoltzmDist}, thermodynamic equilibrium in the computer would lead to:
\begin{equation}
    \mathbb{P}(\bm{S}) = \tfrac{1}{Z}e^{-\alpha(\bm{h},\bm{J},\bm{S}) E_{Ising}(\bm{S})}  \label{eq:alphadist}
\end{equation}
where $\alpha(\bm{h},\bm{J},\bm{x}) = \frac{E_0(\bm{h},\bm{J},\bm{x})}{k_BT}$. For $\alpha$ is a dimensionless scaling factor, we will refer to this phenomenon as Hamiltonian scaling.

\smallskip

\paragraph{Estimating $\alpha$} \label{esta}
\noindent While the dependence of $\alpha$ to $\bm{x}$ must be considered in all rigor, it is commonly neglected in the litterature.
The dependence of $\alpha$ on the target Ising energy, i.e. on ($\bm{h},\bm{J}$), can be understood as $\bm{h}$ and 	$\bm{J}$ can be arbitrarily large parameters. Hence, the maximal value of $\Eising(\bm{x})$ can be arbitrarily large as well; as it is physically impossible to get the system of qubits to have an arbitrarily large energy, $E_0(\bm{h},\bm{J})$ must be able to become arbitrarily small, and so does $\alpha(\bm{h},\bm{J}$). Therefore, the order of magnitude of $\alpha$ is nontrivial to estimate.
Estimating the other parameters involved in $\alpha$ (qubit energies and physical temperature) remains nontrivial. %
An insight on the difficulties related to temperature estimation was given by D-Wave~\cite{raymond2016global}.

\smallskip

\paragraph{Compensating for $\alpha$}
\noindent Assuming thermal equilibrium to be reached, compensating for $\alpha$ is still a nontrivial task.  Nelson et al.~\cite{nelson2021high} retrofit an Ising Hamiltonian to get a Boltzmann distribution. They consider processes in which they aim to sample according to a target distribution:
\begin{equation*}
 \mathbb{P}_{a_{in}}(\bm{x})=\tfrac{1}{Z} e^{-a_{in}H(\bm{x})}, \  Z=\textstyle\sum_{\bm{x}' } e^{-a_{in}}H(\bm{x}')
\end{equation*}
 Upon sampling and thus getting a sample set $\mathcal{S}$, they then numerically determine the parameter:
\begin{equation}
\alpha_{out} = \text{argmin} ( a \mapsto \text{dist} (\mathcal{S}, \mathbb{P}_{a} )) \label{eq:defaout}
\end{equation}
with $\mathbb{P}_a$ is defined as $\mathbb{P}_{a_{in}}$,
which provides the best fit of $\mathcal{S}$ by a Boltzmann distribution associated to an Ising energy that is proportional to the target energy. What the authors show is that even in the case where the fit is actually quite good (which corresponds to very limited situations only), inverting the relation between $a_{out}$ and $a_{in}$ seems intractable. Hence, as using in our notation $\alpha=a_{out}/a_{in}$, it appears that compensating for $\alpha$ is a hard task that lies in the way of being able to sample according to a given distribution.

\subsection{Challenge \#5. Limitations related to sampleset sizes} \label{sssiz}

\noindent The maximal size of a sampleset outputted by a D-Wave computer \cite{headquarters2020technical} is $10^4$, which is fine for the optimization task these computers are meant to tackle. For sampling considerations, this may be sufficient 
for a statistic on the whole distribution with a reasonable precision
, but not for a fine representation of the $2^N$ possible states or a statistic with a precision above the CLT convergence rate.
%
Unfortunately, concatenating multiple samplesets isn't a solution due to spin-bath polarization effects, as the first samplesets will bias the following ones.
Hence, trying to reconstruct the target distribution directly from population estimation is very unlikely to yield concrete results. As estimating the target distribution through population estimation would require some $100\times2^N$ samples, an exponential complexity arises, questioning the actual speedup obtained with the quantum method compared to classical methods such as Markov-Chain Monte-Carlo algorithms \cite{hinton2002training} (as the main difference between both would be the physical time to draw a sample).

\smallskip

\textit{An attempt to address the issue and its limitation}

\noindent As a D-Wave computer outputs not only states, i.e. the $\bm{x}$, but also their associated energies $\Eising(\bm{x})$, a possible solution to our problem would be to estimate the energies of as many states as possible through sampling, manually calculate a Boltzmann distribution on the discovered states, and arbitrarily give the undiscovered states a probability of 0. Unlike a bruteforce calculation, no exponential runtime would be involved because here calculating $Z$ involves at most $10^4$ terms and not an exponential number. Now,

\begin{itemize}[wide]
    \item If there are less that $10^4$ possible states and they are all discovered: this method would yield the target Boltzmann distribution up to errors on the energies, at the same computational cost as classical bruteforce methods.
    \item If there are more than $10^4$ possible states: this method could still be relevant provided that all the states with a non-negligible probability are discovered and that the sum of the probabilities of the undiscovered states is very small. While a rigorous mathematical estimation of the distance between the estimated and the target distributions remains to be done, the physical intuition allows one to think that those distribution would be close enough for many practical applications.
    \item If there are many more than $10^4$ possible states: sampling  according to the target distribution may simply give $10^4$ different states. 
    There, we will assign  a bigger probability to states with a lower energy, thus reducing the distance with the target distribution in comparison of performing a simple population estimation and assigning to each sampled state a probability of 1/$10^4$.
\end{itemize}

\smallskip

\noindent Trying to implement this method, we encountered two major drawbacks. The first one comes from the design of D-Wave computers to be efficient solvers: even on extremely small examples (3 qubits), not all states are discovered in $10^4$ samples. The second one is unpredictable Hamiltonian scaling, as explained in \Cref{Hscal}.


%% file: conclusion.tex
\section*{Conclusion}

\noindent Although it has been successfully tested in the literature on specific use cases, using D-Wave QC to speed up the sampling of general Boltzmann distributions is still an open question. We described five challenges that must be  overcome  to  performing  such  sampling.  It  still remains to  find  out  a  protocol  for  such  sampling  on  the  current generation  of D-Wave  computers,  or  to  prove  that  such sampling is not possible in the general case.